\documentstyle{article}

\begin{document}

\title{$Z_3$-graded Grassmann Variables, Parafermions and their 
Coherent
 States}
  \author{}

\date{}

\maketitle

\center{\bf M. EL BAZ}\footnote{E-mail address: moreagl@yahoo.co.uk}

\center{\it Facult\'e des sciences, D\'epartement de Physique, 
LPT-ICAC,
 \linebreak Av. Ibn Battouta, B.P. 1014, Agdal, Rabat, Morocco}

\vspace{0.5cm}

\center{\bf Y. HASSOUNI}\footnote{E-mail address: Y-hassou@fsr.ac.ma}

\center{\it Facult\'e des sciences, D\'epartement de Physique,
 LPT-ICAC,
\linebreak Av. Ibn Battouta, B.P. 1014, Agdal, Rabat, Morocco 
\linebreak
 and
\linebreak the Abdus Salam International Centre for Theoretical
 Physics
\linebreak Strada Costiera 11 , 34100 Trieste, Italy}
 
\vspace{0.5cm}

\center{\bf F. MADOURI}\footnote{ E-mail address:
 Fethi.Madouri@ipeit.rnu.tn}

\center{\it Facult\'e des sciences, D\'epartement de Physique, 
LPT-ICAC,
 \linebreak Av. Ibn Battouta, B.P. 1014, Agdal, Rabat, Morocco 
\linebreak
 and \linebreak I.P.E.I.T, 1008 Monfleury, Tunis, Tunisia}

\vspace{1.2cm}

\abstract{A relation between the $Z_3$-graded Grassmann variables and parafermions is established. Coherent states are constructed as a direct consequence of such a relationship. We also give the analog of the Bargmann-Fock representation in terms of these Grassmann variables.} 
 
\vspace{2cm}
\pagebreak

\section{Introduction}

 Recently Kerner \cite{kerner93, kerner96, kerner00} investigated the 
use
 of $Z_3$-graded structures instead of $Z_2$-graded ones in physics. 
This
 leads to some interesting results especially when gauge theories are constructed 
using
 these structures.

One of these structures is the $Z_3$-graded analog of the Grassmann algebra. The generators of this algebra obey some ternary relations instead of the usual binary relations (anticommutation) for the generators of the conventional Grassmann algebra ($Z_2$-graded one).

 In this letter we relate the $Z_3$-graded Grassmann variables to the
 parafermionic harmonic oscillator. We will use this result in the 
subsequent
 construction of coherent states related to this parafermionic 
oscillator. The construction of these Coherent states as well as all the subsequent derivations are carried out in the spirit of what was achieved in the case of the fermionic harmonic oscillator, {\it via} the $Z_2$-graded Grassmann variables \cite{ohnuki}. This will prove useful in understanding the physical meaning of such
 structures.

 We start with a review of the $Z_3$-graded Grassmann algebra and its
 properties. In section 3 we review the parafermionic harmonic 
oscillator
 algebra. In section 4 we discuss the relationship between the 
Grassmann
 algebra of section 2 and the harmonic oscillator of section 3. This 
will
 allow us in section 5 to construct the coherent states for this system
 in a clear and unambiguous manner. In this construction we propose a new form for the resolution of the identity operator, different from the usual one. This will permits in section 6 to give an analog of the Bargmann-Fock representation space: The space of Grassmann representatives.

In section 7 we suggest how the supersymmetric formulation of this construction should look like.

\section{ $Z_3$-graded Grassmann algebra}

 $Z_3$ is the cyclic group of three elements. It can be represented in 
the
 complex plane as multiplication by the primary cubic root of unity 
 $ q=e^{2\pi i \over 3}$, $q^2$ and $q^3=1$. The analog of the
 $Z_2$-graded Grassmann algebra can be introduced as follows:

 Consider an associative algebra spanned by $N$ generators
 $ \xi_a;\;\; a =0,= 1,...., N$, between which only ternary relations
 exist. By this we mean that the binary products of any couple of such
 elements are considered as independent entities, i.e.  $\xi_a\xi_b$ are
 independent of $\xi_b\xi_a$ where $a, b = 0, 1..., N$).
 Instead, the analog of the anti-commutation in the $Z_2$-graded
 case is given by the following ternary relations:

\begin{equation} 
\xi_a\xi_b\xi_c = q\xi_b\xi_c\xi_a = q^2 \xi_c\xi_a\xi_b \;\;\;\;\;\;
 a,b,c = 0,1,...,N
\end{equation}

Two important properties follow automatically:
\begin{itemize}
\item{ The third power (or higher) of any generator vanishes :

\begin{equation}
(\xi_a)^3 = 0
\end{equation}}

\item{ Any product of four or more generators also vanishes :

\begin{equation}
\xi_a\xi_b\xi_c\xi_d = 0 \;\;\;\;\;\; a,b,c,d = 0,1,...,N
\end{equation} }

\end{itemize}

 At this level one remarks that there is no symmetry between the 
grade-1
 elements (the $\xi$'s) and the grade-2 elements
 (the $\xi\xi$'s)\footnote{ The algebra contains $N^2$ grade-2 
elements,
 while it contains only $N$ grade-1 elements}.
 It seems normal that these elements should play a symmetric role with
 regard to $q$ and $q^2$. This symmetry is restored in the most natural 
way,
 by adding $N$ grade-2 generators $\bar{\xi}_a$ (the duals of
 $\xi_a$)\footnote{ its dual, in the sense that the product of a $\xi$ 
and a
 $\bar{\xi}$ is a grade-0 element.i.e it behaves like a scalar}. These 
extra
 elements satisfy the same relations as the $\xi$'s, but with $q$ 
replaced
 by $q^2 :$

\begin{equation}
\bar{\xi}_a\bar{\xi}_b\bar{\xi}_c =
 q^2\bar{\xi}_b\bar{\xi}_c\bar{\xi}_a
\end{equation}
and their binary products with the $\xi$'s satisfy :

\begin{equation}
\xi_a \bar{\xi}_b = q \bar{\xi}_b \xi_a
\end{equation}

 Now, by requiring that the grades of factors add up (modulo 3) under 
 multiplication and that grade-0 elements commute with all other 
elements,
 and that grade-1 with grade-2 elements satisfy the same relation as in 
(5),
 then many additional terms must vanish ( terms like 
$\xi_a\xi_b\bar{\xi}_c$
 for example ). The resulting algebra contains only the elements of the 
form:

\begin{item} 
\item{ Grade-0 \hspace{1cm} : \hspace{1cm} $ I, \;\; \xi \bar{\xi},\;\;
 \xi\xi\xi,\;\; \bar{\xi}\bar{\xi}\bar{\xi}$ }

\item{ Grade-1 \hspace{1cm} : \hspace{1cm} $ \xi, \;\; 
\bar{\xi}\bar{\xi} $ }

\item{ Grade-2 \hspace{1cm} : \hspace{1cm} $ \bar{\xi}, \;\; \xi \xi $ 
}

\end{item}
and its dimension is: $\displaystyle D= { 3 + 4N + 9N^2 + 2N^3 \over 3} 
$

 The case $N=1$ was considered in detail in \cite{chung}, however the
 author imposeded there also binary relations between elements of the 
same
 grade, and the arguments for this are not too convincing. In the 
following
 sections we are going to investigate  the same case, however, without
 imposing such binary relations. This will be more consistent with what 
we
 announced previously based on Kerner's works \cite{kerner93, kerner96,
 kerner00}

\section{ Parafermionic Harmonic Oscillator}

 The most natural way to introduce a parafermionic harmonic oscillator 
which
 can be related to the $Z_3$-graded Grassmann algebra is through the 
deformed
 harmonic oscillator algebra generated by the operators ${ a, a^+, N}$.
 These operators shall satisfy the following commutation relations:

\begin{eqnarray}
aa^+ - qa^+a &=& q^{-N} \nonumber \\
Na - aN &=& -a \nonumber \\
Na^+ - a^+N &=& a^+ \\
q^N a^+  &=&  a^+q^{N+1} \nonumber \\
q^N a  &=&  a q^{N-1} \nonumber
\end{eqnarray}
where $q$ is an arbitrary complex parameter of deformation.

 Now, when $q$ is the primitive $k^{th}$-root of unity (i.e. $q=e^{2\pi 
i
 \over k}$), one can prove that the annihilation and creation operators 
$a$
 and $a^+$ are nilpotent of degree $k$ \cite{mansour}\footnote{This 
means
 that no more than $k-1$ parafermions are allowed to occupy the same 
state.
 This is a straightforward generalization of the Pauli exclusion 
principle.}

\begin{equation}
(a)^k = 0 \;\;,\;\; (a^+)^k = 0
\end{equation}

 A comparison of the above relation with (2) suggests that $Z_3$-graded
 Grassmann variables in section 2 can be used as a representation of 
this
 oscillator (for $k= 3$). This is indeed the case, as we shall see in
 the next section.

 The Fock space representation of this algebra is given by:

\begin{eqnarray}
a|n> & = & \sqrt {[n]} |n-1> \nonumber \\
a^+|n> & = & \sqrt {[n+1]} |n+1>  \\
N |n> & = & n |n> \nonumber
\end{eqnarray}
where ${|n>; n= 0,1,...,k}$ is the usual Fock space orthonormal basis, 
and

\begin{equation}
[n] = { q^n - q^{-n} \over q - q^{-1}}
\end{equation}

 At this point we should stress the fact that for the case we are 
considering
 ($q$= root of unity) $a^+$ is still the usual hermitian conjugate of 
$a$.
 This reflects itself in the representation (8), where $ [n] = 
\bar{[n]}$
 which is not the case for a generic $q$. One then has to introduce two 
other
 operators in the algebra, a creation (hermitian conjugate to $a$) and 
an
 annihilation operator (hermitian conjugate to $a^+$), but we do not 
intend
 to consider this possibility here\footnote{of course we should suppose 
that 
 the operators $a$ and $a^+$ obey, in addition to (6), the same 
equations but 
 with $\bar q$ instead of $q$, for instance $ aa^+ - \bar q a^+a = 
q^N$}.

\section{ $Z_3$-graded Grassmann variables and the parafermionic 
harmonic
 oscillator:}

 Before we start to construct the coherent states associated to the
 $Z_3$-graded harmonic oscillator we should define the relationship 
between
 the $Z_3$ grassmannian variables introduced in section 2 and the $Z_3$
 harmonic oscillator's operators introduced in section 3.

 According to \cite{chung}: $|0>$ behaves like a grade-0, $|1>$ like a
 grade-2 element, while $|2>$ behaves like a grade-1 element. This, 
combined
 with the equations (8), defines the algebra completely :

\begin{eqnarray}
a^+|0> &=& |1> \nonumber \\
a|1> &=& |0>
\end{eqnarray}
or
\begin{eqnarray}
a^+a^+|0> &=& \sqrt{[2]}|2> \nonumber \\
a|2> &=& \sqrt{[2]}|1>
\end{eqnarray}

 This fact suggests to interpret $a^+$ as being of grade-2 and $a$ as 
being
 of grade-1.

 Now, using the conventions cited at the end of section 2, one is led 
to the
 following relations:

\begin{eqnarray}
\xi a^+ &=& q a^+ \xi \nonumber \\
\bar {\xi} a &=& \bar q a \bar {\xi}
\end{eqnarray}

 To be consistent with what we announced in section 2, no relations are
 imposed on the products $\xi a$ and $a\xi $ (the same stands for 
$\bar{\xi}
 a^+$ and $a^+ \bar{\xi}$). Furthermore no analogs of the ternary 
relation
 (1) can be imposed on products of the form $a\xi a^+$; this is due to 
the
 non-conventional commutation relations in (6). In order to evaluate 
such 
 products, one should use the relations (6) and (12).

 We notice that in \cite{chung} the author chose $\xi$ and $\bar{\xi}$ 
to
 commute with $a$ and $a^+$. This does not seem consistent, since even 
in
 the fermionic case (i.e. $Z_2$ case) that we are supposed to 
generalize,
 the $\xi$'s do anti-commute with the $a$'s !

\section{Coherent States}

 In what follows, we shall drop the rule cited at the end of section 2 
(the
 grades adding up mod-3 under multiplication, the grade-1 elements 
behaving 
 as the $\xi$'s, grade-2 elements as the $\bar \xi$'s and grade-0 
elements as
 scalars.). As a matter of fact, it is interesting to note that unlike 
in the
 $Z_2$-graded case, where such a rule follows automatically\footnote{In 
fact
 in this case, the anticommutation of the generators of the algebra and 
the
 associativity, imply automatically that all grade-0 elements commute 
with
 all the other elements (scalar behavior), and the grade-1 elements
 anti-commute with each other ($\xi$'s behavior)}, in the $Z_3$-graded
 case this is no more true, so that this rule have to be explicitly 
imposed.

 We can now proceed towards the construction of coherent states:

 Considering the possibilities that are allowed in order to construct 
 coherent states for the $Z_3$-graded harmonic oscillator introduced in 
 section 3, it is easy to see that the only consistent combination is 
given by:

\begin{eqnarray}
|\xi> &=& f(a^+\xi) |0> \nonumber \\
&=& |0> + a^+\xi \;|0> - a^+\xi a^+\xi \;|0>
\end{eqnarray}
 where $f(a^+\xi) = 1 + a^+ \xi - a^+ \xi a^+ \xi$ generalizes the
 function ($ 1 + a^+\xi $) in the fermionic case \cite{ohnuki}.

 Using (12) this state can be rewritten as:

\begin{equation}
| \xi > = |0> + q^2 \xi |1> - \sqrt{[2]}\xi^2|2>
\end{equation}

 In what follows, we shall demonstrate that these states satisfy the 
usual
 coherence criteria and therefore can represent genuine coherent states 
for 
 the parafermionic harmonic oscillator.

 First of all, using (6), (8), (12) and the rules cited at the end of 
section
 4, it's easy to see that the states (13) are indeed eigenstates of the
 annihilation operator

\begin{equation}
a|\xi > = \xi |\xi >
\end{equation}

 One can also compute the scalar product of two such states using the 
same
 relations and the orthonormality of the Fock space basis, the result 
being
 then as follows :

\begin{equation}
<\bar {\xi_1}|\xi_2> = 1 + q^2 \bar {\xi_1}\xi_2 -
 q\bar{\xi_1}\xi_2\bar{\xi_1}\xi_2 = g(\bar{\xi_1}\xi_2) 
\nonumber
\end{equation}
where
\begin{equation}
<\bar {\xi_1}| = <0| + q<1| \bar {\xi_1} - \sqrt{[2]} <2|\bar 
{\xi_1}^2
\end{equation}

 A resolution of the identity is also possible in terms of the states 
(13) or
 alternatively (14). In fact, since the three eigenvectors $|0>$, $|1>$ 
and 
 $|2>$ form an orthonormal basis, the identity operator may be 
expressed as

\begin{equation}
I = |0><0| + |1><1| + |2><2|
\end{equation}
 and using the integrals defined by Majid \cite{majid}:

\begin{eqnarray}
\int 1 \, d\xi  &=& \int \xi \; d \xi = 0 \nonumber \\
\int 1 \, d \bar{\xi}  &=& \int  \bar{\xi} \; d \bar{\xi } = 0 \\
\int  \xi^2 \, d \xi &=& \int \bar {\xi}^2  d\bar {\xi}^2 = 1
\nonumber
\end{eqnarray}
 then the identity operator, in terms of the $|\xi>$'s is given by:

\begin{equation}
\int d\bar{\xi} \; d\xi \; w(\bar{\xi}\xi) \; |\xi><\bar{\xi}| = I
\end{equation}

 where the weight function is defined as :

\begin{equation}
w(\bar{\xi}\xi) = -q + \bar{\xi}\xi + \bar{\xi}\xi\bar{\xi}\xi
\end{equation}

 This completes our proof of the fact that the states (13) are indeed
 coherent states\cite{klauder}.

 What we feel particularly not comfortable with in this construction is the fact that the different functions involved (in the process of building up these coherent states), does not seem to generalize the usual exponential function which permits the construction of fermionic coherent states. In fact none of the functions $f$, $g$ or $w$ resemble to any known deformation or generalization of the exponential function.

Nevertheless, and apart from aesthetic reasons, this will not constrain us from pushing further the analogy between these states and the conventional fermionic coherent states. Indeed, in addition to properties (15),(16) and (20) (which constitute the essential defining properties of a coherent state), we shall show in the next section that these states provide us with a Bargmann-Fock representation analog. To do this we have to express the identity operator; using  states (14); in a very special manner:

\begin{equation}
I = \int | \xi > \; d\bar \xi d\xi \; w(\bar \xi \xi) < \bar \xi|
\end{equation}

The choice of this form will become clear in the next section.

This resolution of the identity is not equivalent to the one given in (20). In contrast to the fermionic (or even bosonic) case where this two resolutions are equivalent, so it is sufficient to give one form to recover the other.

Note that the weight function $w$, involved in the two forms (20) and (22), is the same.

For completeness we shall also define the behavior of $d\xi$ and $d\bar \xi$. Indeed, in order to be consistent with what precedes we can prove that $d\xi$ should behave like the $\xi$, and $d\bar \xi$ like the $\bar \xi$:
\[
d\xi \; \bar \xi = q \bar \xi \; d\xi \;\;\;\; , \;\;\;\; \xi \; d\bar \xi = q d\bar \xi \; \xi \;\;\;\; , \;\;\;\; d\xi \; d\bar \xi = q d\bar \xi \; d\xi \; \dots
\]

\section{Grassmann representatives of state vectors}

One of the most interesting features of coherent states is that they permit the construction of the so called Bargmann-Fock representation space. In this section we shall prove that the states constructed (13) do also exhibit this feature and one can construct an analog of this representation space.

In analogy with the fermionic case, for any state vector $|\psi >$ in the Fock space, we can now define its Grassmann representative by:

\begin{equation}
\psi (\bar \xi ) = < \bar \xi | \psi>
\end{equation}
and its adjoint:
\begin{equation}
\bar \psi (\xi ) = < \psi | \xi>
\end{equation}

Note that the Grassmann representatives defined in this manner depend on the Grassmann variable $\bar \xi$. This is just a matter of convention, we could have defined the coherent states in (13) using $\bar \xi$ instead of $\xi$, the Grassmann representatives will then be expressed in terms of the $\xi$'s.

There is a one to one correspondence between the space of Grassmann representatives and the Fock space. In fact a state $|\psi >$can be uniquely defined when its representative $\psi (\bar \xi )$ is given.

Now, we shall use the resolution of the identity (22) to define the inner product in this space (of Grassmann representatives). We procede as in the usual case, i.e. sandwiching eq(22) between two states $<\psi |$ and $|\phi >$:

\begin{equation}
<\psi | \phi > = \int \bar \psi(\xi )\;d\bar \xi \; d\xi \; w(\bar \xi \xi ) \; \phi(\xi )
\end{equation}

Taking this equation as a definition for the inner product in the space, permits to describe this representation space as an analog of the Bargmann-Fock representation space. The role of the weight function $exp(\bar z z)$ in this last one is played, in our case by $w(\bar \xi \xi)$ given in (21).

Next we demonstrate that the scalar product (25) yield the same result as the conventional scalar product in the Fock space.

The basis vectors $|n>$ are represented in this Bargmann-Fock space analog by:

\begin{equation}
<\bar \xi | n> = c_n {\bar \xi }^n \;\;\;\;\;\;\; c_n: \hbox{a complex parameter}
\end{equation}
as a matter of fact, the orthonormality of this basis should be translated in this representation:
\begin{equation}
<n|m> = c_n^*c_m \int \xi ^n \; d\bar \xi d\xi \; w(\bar \xi \xi)\; {\bar \xi}^n = \delta _{n,m}
\end{equation}

We show this for two representative cases:

$<0|0> = 1$:

we have
\[
<0|\xi > = <\bar \xi |0> = 1
\]
\begin{eqnarray}
<0|0> &=& \int d\bar \xi d\xi \; w(\bar \xi \xi) = \int d\bar \xi d\xi (-q + \bar \xi \xi + \bar \xi \xi \bar \xi \xi ) \nonumber \\
&=& \int d\bar \xi d\xi \; \bar \xi \xi \bar \xi \xi = \int d\bar \xi d\xi \; \xi ^2 {\bar \xi}^2 = 1
\end{eqnarray}

$<0|1> = 0$
\[
<\xi |1> = q <1|\bar \xi |1> = \bar q \bar \xi
\]
\begin{eqnarray}
<0|1> &=& \bar q \int d\bar \xi d\xi \; w(\bar \xi \xi ) \; \bar \xi \nonumber \\
&=& \bar q \int d\bar \xi d\xi \; (-q + \bar \xi \xi + \bar \xi \xi \bar \xi \xi ) \bar \xi = 0
\end{eqnarray}

the calculations are done in the same way for the other examples.

\section{$Z_3$-graded supersymmetric coherent states}

 In addition to the parafermionic harmonic oscillator related to the
 $Z_3$-Grassmann algebra, we need a bosonic harmonic oscillator, in 
order to
 construct a $Z_3$-graded supersymmetric one.

 The bosonic harmonic oscillator algebra is generated by the triplet 
 $\{M,b,b^+\}$, satisfying the usual commutation relations. 

\[
[b,b^+] = \, 1 \;\;\;, \;\;\; [b,M] = - b \;\;\; , \;\;\; [b^+,M] = \, 
b^+
\]
acting on the usual Fock space basis $\{ |m>, m= 0,1,...\}$ as:

\[ 
b|m> = m |m> \;\;\;, \;\;\; b^+|m> = (m+1)|m+1> \;\;\;, \;\;\;M|m> =
 m|m>
\]

The bosonic coherent states are given for a complex $z$ by:

\begin{equation}
|z> = \sum _{m=0}^{\infty} {z^m \over {\sqrt{m!}}} \, |m>
\end{equation}

 As in the case of $Z_2$-graded supersymmetry, a $Z_3$-graded 
supersymmetric
 coherent state is obtained by coupling this states with the states 
(13):

\begin{eqnarray}
|z,\xi > &=& |z> \otimes \; |\xi > \\
&=& D(z,\xi )\; |0>\otimes \; |0 > \nonumber
\end{eqnarray}
 where $D(z,\xi) = e^{zb^+}f(\xi a^+)$

\vspace{0.4cm}

 We believe that the results obtained in this paper constitute a 
non-trivial
 physical application of the $Z_3$-graded structures. As a matter of 
fact, 
 taking the view according to which the $Z_3$-graduation is a natural
 generalization of conventional non-commutative geometries
 \cite{kerner96, kerner00}, we can use the above results to construct 
the
 generalized quantum plane associated to it, then investigate its
 properties and try to apply them to physical problems.
 Another area where we can use these results is of course 
supersymmetry.
 A deeper investigation of the states defined in section 6, in the 
light of 
 what has been achieved in \cite{kerner92}, remains to be performed.

As for the Bargmann-Fock representation analog, this will prove useful to construct a path integral approach in terms of the Grassmann variables for the description of this parafermions.

 The results concerning these points will be soon available in the
 forthcoming study.

 \section*{Acknowledgments}

 Two of the authors (M.EB) and (Y.H) wish to express their thanks for
 the hospitality extended to them by the Abdus Salam International 
Centre for Theoretical Physics (Trieste, Italy) where part of the work was achieved.

\end{document}